\documentclass[12pt,twoside,fleqn]{article}
\usepackage{espcrc1}
\usepackage{graphicx}

\newcommand{\AmS}{{\protect\the\textfont2
  A\kern-.1667em\lower.5ex\hbox{M}\kern-.125emS}}

\title{Hyperon semileptonic decay constants and quark-spin content
of the octet baryons}

\author{\underline{Hyun-Chul Kim}\address{Department of Physics,
Pusan National University,\\ 609-735 Pusan, Republic of Korea}, %
Micha{\l} Prasza{\l}owicz \address{Institute of Physics,
Jagellonian University, \\
Cracow, Poland}
and Klaus Goeke \address{Institute for Theoretical Physics II,
Ruhr-University Bochum, \\ D-44780 Bochum, Germany }
} %
\date{\today}

\begin{document}

\maketitle
\vskip -3.5in
\rightline{\hbox{PNU-NTG-04/2000}}
\vskip3.3in
\begin{abstract}
We present the recent investigation on the effect of
the SU(3) symmetry breaking in the spin structure of the octet
baryons.  
\end{abstract}

\vspace{1.2cm}
 The spin structure of the nucleon has been a hot
issue since the measurement of the first moment of the proton spin
structure function $g_{1}^{p}$~\cite{EMC} by the European Muon
Collaboration (EMC).  The EMC result indicates that the strange
quark is polarized in the opposite direction to other light quarks.
It implies that the spin of the nucleon is not carried by quarks only.  
 While it is now known that a large fraction of the nucleon spin is
provided by gluons and their orbital angular momenta, it is still
very important to understand the mechanism of how the quarks carry
the nucleon spin. The information on the nucleon spin carried by
quarks comes from two different sources: the first moment of the
spin structure function $g_{1}^{{\rm {p}}}(x)$ of the proton and
the experimental data of hyperon semileptonic decay (HSD).
With SU(3) symmetry assumed, one obtains two reduced matrix
elements $F$ and $D$. Taking for $F=0.46$ and for $D=0.80$, one
gets for the proton: $\Delta u_{{\rm {p}}}=0.79$, $\Delta d_{{\rm
{p}}}=-0.47$ and $\Delta s_{\rm p}=-0.13$, which implies $\Delta
\Sigma _{\rm p}=0.19$, quite a small number as compared with the
naive expectation from the quark model: $\Delta \Sigma _{{\rm
{p}}}=1$.  

Recently, the spin content of the $\Lambda$ hyperon has drawn attention to us.
In contrast to the nucleon, the nonrelativistic quark model predicts
that the spin of the $\Lambda$ comes solely from the strange quark,
since the up and down quarks constitute the spin singlet.
Furthermore, the $\Delta q_\Lambda$ is deeply related to 
the fragmentation funtion which can be measured 
experimentally~\cite{BurkardtJaffe}.    

However, note that $\Delta \Sigma _{{\rm {p}}}$ and 
$\Delta \Sigma_\Lambda$ are {\em not
directly measured}.  The standard way to extract $\Delta \Sigma
_{\rm B}$ is to assume the SU(3) symmetry for HSD. In this
case it is enough to take {\em any} two decays and data of deep
inelastic scattering as input.  With SU(3) symmetry assumed, one
can pick out two different values of the constants of HSD
among the existing data.  One normally uses neutron and $\Sigma^-$
$\beta$ decays as an input. However, if the SU(3) symmetry
breaking were not important, any pair out of the six known HSD
constants would give roughly the same number for $\Delta \Sigma
_{\rm B}$. This is, however, {\em not} the case.  In fact, 
the effect of SU(3) symmetry breaking in HSD is not at all 
small.  Furthermore, the uncertainty of the HSD 
constant for $\Xi^{-}\rightarrow \Sigma^0$ is noticeably larger than 
those for the other processes.  Thus, it is still of great significance to
understand the effect of SU(3) symmetry breaking on the quark-spin
content of the nucleon.

In order to analyze the symmetry breaking in weak decays, it is
inevitable to resort to some specific model~\cite{KarLip}. We
employ the chiral quark-soliton model($\chi$QSM): See
Ref.\cite{review} and references therein) to incorporate the SU(3)
symmetry breaking into the semileptonic decays. The $\chi$QSM
satisfactorily describes the axial-vector properties of
hyperons (see Ref.~\cite{KimPoPraGo} and references therein.).
Since the symmetry breaking pattern of the $\chi$QSM is identical
to the one derived in large $N_{\rm c}$ QCD \cite{Man}, our
analysis is in fact much more general than the model itself.
In addition, we employ the experimental data for HSD in order 
to fix the dynamical variables in the model, so that we can minimize
the model-dependence of the present study.  

However, since $g_{{\rm A}}^{(0)}($B$)$ does not correspond to the
SU(3) octet axial-vector current, a further assumption is required
to connect it to the octet axial-vector constants.  The $\chi$QSM
(as most of the hedgehog models) has a remarkable virtue of
connecting the $g_{{\rm A}}^{(0)}($B$)$ with $g_{{\rm A}}^{(3)}$
and $g_{{\rm A}}^{(8)}$, and the semileptonic decay constants in a
direct manner.  This connection introduces a model dependence into
our analysis. However, there is no significant numerical
difference between the results obtained with and without this
model dependent ingredient~\cite{KimPraGo2}.  Thus, we would like
to report in brief the recent investigation on the effect of the
SU(3) symmetry breaking, based on
Refs.~\cite{KimPraGo2,KimPraGo3}, to which we refer for details.

Figure 1 shows the dependence of the integrated quark densities 
$\Delta q_{\rm n}$ of the nucleon on the strange quark mass $m_{\rm
s}$, of which definitions can be found in Ref.~\cite{KimPraGo2}:

\vspace{0.4cm}
\begin{center}
\includegraphics[width=7cm,height=5.5cm]{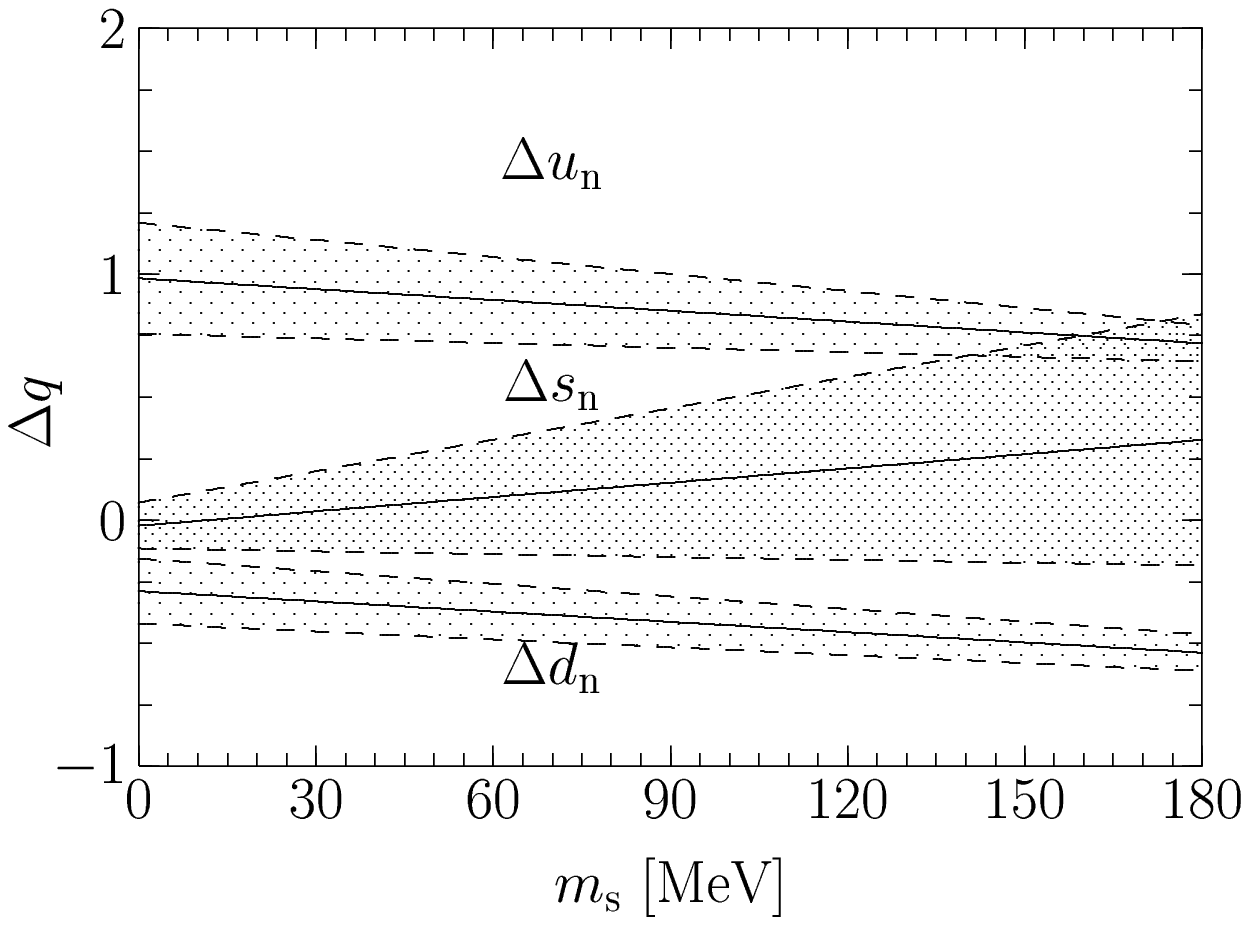}
\vspace{0.2cm}

\parbox{14.0cm}{\footnotesize{\bf Figure 1}: 
The dependence of the $\Delta q_{\rm n}$ on the strange quark mass 
$m_{\rm s}$.}
\end{center}

\vspace{0.3cm} 
While the up and down quark densities have smaller errors as 
$m_{\rm s}$ increases, the uncertainty of the $\Delta s_{\rm n}$ 
becomes much larger.  With $m_s = 180$ MeV, 
we obtain $\Delta s_{\rm n} = 0.33\pm 0.51$.  It indicates that
$\Delta s_{\rm n}$ is in practice undetermined.  The reason is that 
the error of the semileptonic decay constant for the process 
$\Xi^{-}\rightarrow\Sigma^0$ propagates strongly in the $\Delta s_{\rm n}$.

Figure 2 plots the flavor decomposition of the quark-spin content of the 
$\Lambda$.  As in the case of the nucleon, the error of the 
$\Delta s_\Lambda$ is noticeable with $m_{\rm s}$ increased.

\vspace{0.4cm}
\begin{center}
\includegraphics[width=7cm,height=5.5cm]{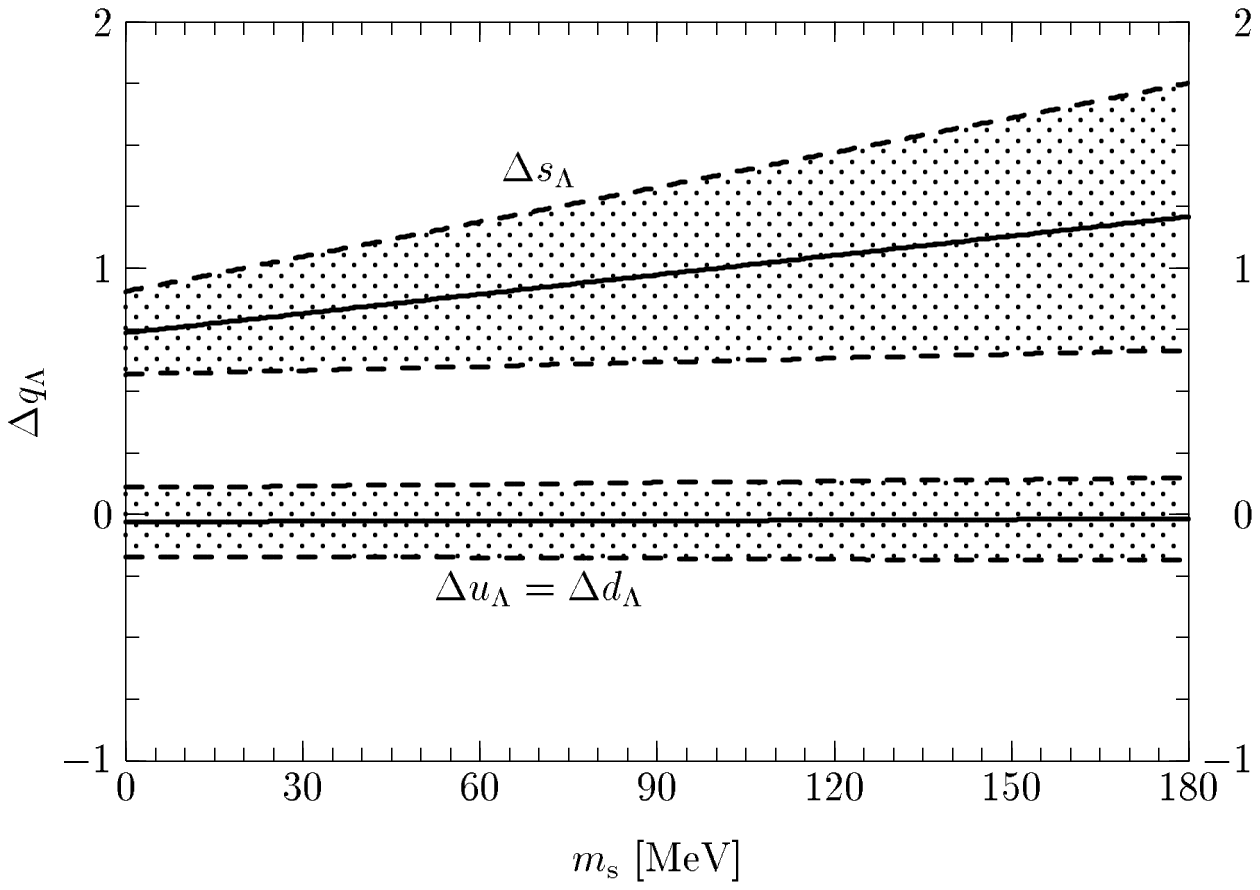}
\vspace{0.2cm}

\parbox{14.0cm}{\footnotesize{\bf Figure 2}: 
The dependence of the $\Delta q_\Lambda$ 
on the strange quark mass $m_{\rm s}$.}
\end{center}

Figure 3 depicts the splitting of the $\Delta\Sigma_{\rm B}$s due
to the strange quark mass.  The result is according to the
following equation:
\begin{equation}\label{ga0}
g_{\rm B}^{(0)}\hat{S}_3\;=\; a_3 \hat{S}_3 + \sqrt{3} m_{\rm s}
(a_5 - a_6) \langle B| D_{83}^{(8)} | B\rangle,
\end{equation}
where $\hat{S}_3$ is the third component of the spin operator. The
$m_{\rm s}$ stands for the strange quark mass which plays a role
of breaking SU(3) symmetry.  $\langle B| D_{83}^{(8)} | B\rangle$
is the matrix element of the SU(3) Wigner function. The $a_i$
denote the dynamical variables from the model which are fixed by
the HSD data.  They appear also in the octet axial-vector
constants, which makes it possible to link the octet axial-vector
constants to the octet one.  

\vspace{0.4cm}
\begin{center}
\includegraphics[width=7cm,height=5.5cm]{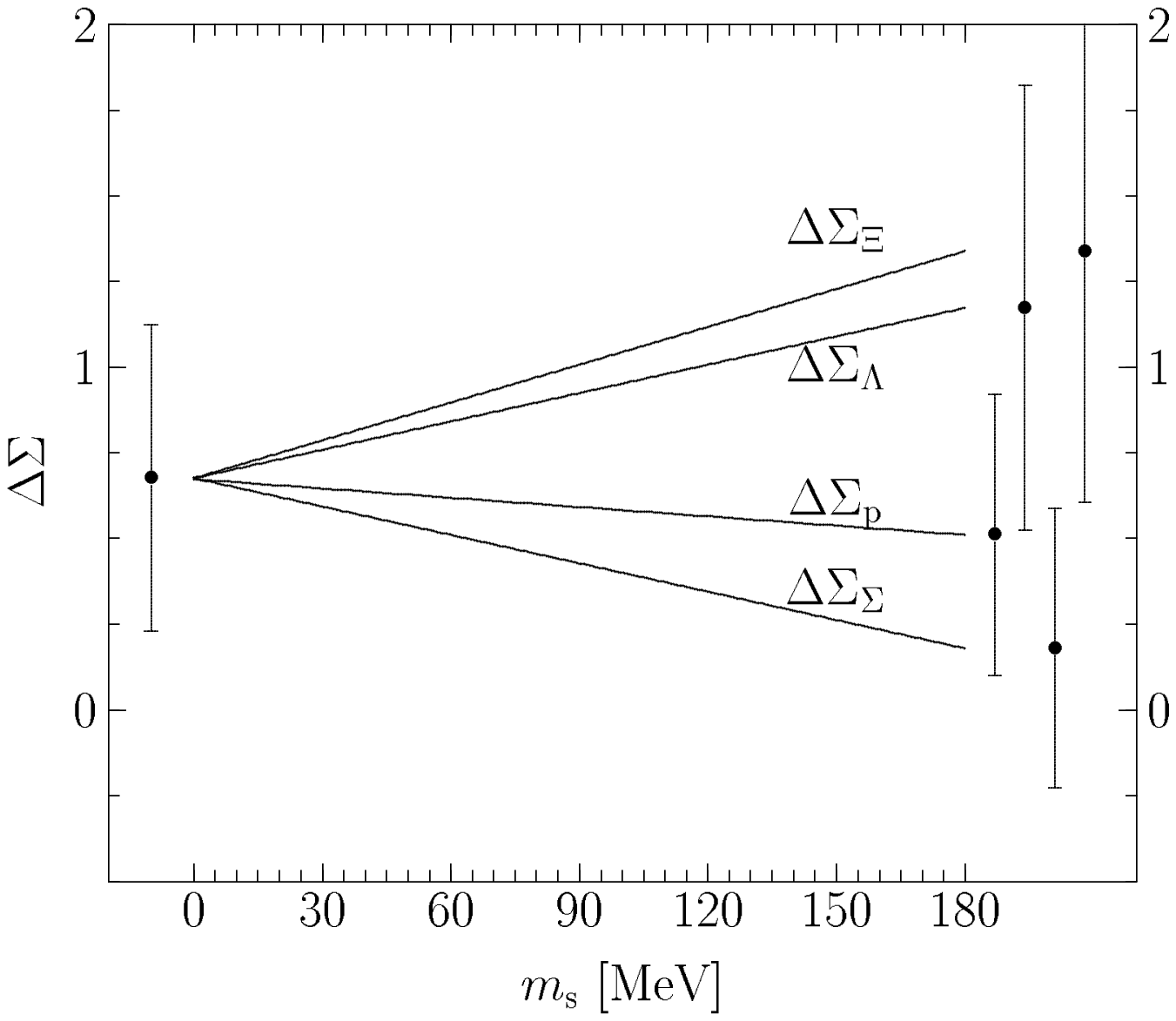}
\vspace{0.2cm}

\parbox{14.0cm}{\footnotesize{\bf Figure 3}:Effect of SU(3) symmetry breaking
on the $\Delta \Sigma _{{\rm {B}}}$.}
\end{center}
Actually, the simplest version of the Sykrme model without vector
mesons gives $a_3 = 0$.  In the case of the $\chi$QSM provides
nonzero $a_3$.  In the limit of the large soliton, the ratio of
$a_3/a_i$ goes to zero, which is in agreement with
Ref.~\cite{BroEllKar}.  When the size of the soliton becomes
closer to zero, which corresponds to the limit of the
nonrelativistic quark model, $g_{\rm p}^{(0)}$ turns out be one, as it
should be.  The removal of the degeneracy in $\Delta\Sigma_{\rm
B}$s comes from the second term in Eq.(\ref{ga0}).  While the
$\Delta\Sigma_{\rm p}$ and $\Delta \Sigma_\Lambda$ show relatively
mild dependence on $m_{\rm s}$, the $\Delta \Sigma_\Xi$ and
$\Delta\Sigma_\Sigma$ depend quite strongly on $m_{\rm s}$.  Large
error bars with $m_{\rm s}$ is due to the propagation of the
uncertainty in $\Xi$ decay to the variables $a_i$.  From figure 3,
we find a very interesting thing.  With the effect of the SU(3)
symmetry breaking, $\Delta \Sigma_\Lambda$ and $\Delta \Sigma_\Xi$
get larger and finally quite closer to one.  It indicates that the
spins of $\Lambda$ and $\Xi$ are almost carried by quarks. That
is, $\Lambda$ and $\Xi$ behave more ``nonrelativistically''.
Hence, extracting the $\Delta\Sigma_\Lambda$ from the experimental
data as precisely as possible and comparing it with that of the nucleon, 
one would see the effect of the SU(3) symmetry breaking, explicitly.

In summary, we have presented in this talk a recent investigation
on the effect of the SU(3) symmetry breaking within the framework
of the chiral quark-soliton model, incorporating the experimental
data of hyperon semileptonic decay.  Because of the large
uncertainty in the data for the $\Xi$ decay, one is not able to
determine the $\Delta s_{{\rm n},\Lambda}$, which means that one can not 
tell for sure about how much the strange quark carries the nucleon spin.

The SU(3) symmetry breaking removes the degeneracy of the $\Delta
\Sigma_{\rm B}$s. Therefore, knowing the difference   
between the $\Delta\Sigma_{\rm p}$ and $\Delta\Sigma_\Lambda$ 
will bring out the effect of the SU(3) symmetry breaking. 

The $\Lambda$ and $\Xi$ behave
more ``nonrelativistically'' than the nucleon.  

In order to determine the spin content of the SU(3) baryons
carried by quarks, it is highly required to reduce the experimental
data of $\Xi$ semileptonic decay. 

\section*{Acknowledgments}
The work of H.C.K has been supported by the Research Fund of 
Pusan National University.  The work of M.P. has been supported by Polish 
KBN Grant PB~2~P03B~019~17. The work of K.G. has been supported by the BMBF,
the DFG, and the COSY--Project(J\"{u}lich).

\end{document}